# Covalent Crosslinking of Porous Poly(Ionic Liquid) Membrane via a Triazine Network

Karoline Täuber,[†] Alessandro Dani,[†] Jiayin Yuan*

Department of Colloid Chemistry, Max Planck Institute of Colloids and Interfaces, Am Mühlenberg 1 OT Golm, D-14476 Potsdam, Germany.

*Supporting Information Placeholder*

**ABSTRACT:** Porous poly(ionic liquid) membranes that were prepared via electrostatic cross-linking were subsequently covalently cross-linked via formation of a 1,3,5-triazine network. The additional covalent cross-links do not affect the pore size and pore size distribution of the membranes and stabilize them towards salt solutions of high ionic strength, enabling the membranes to work in a broader environmental window.

Porous polymer membranes are a field of growing interest both in academia and industry.[1-7] Such membranes are composed of polyelectrolytes, where the charge character of the polymer affords a wide range of applications such as sensing, separation and catalysis.[8-19] From a structural point of view, the porous morphology of the membrane is important, especially considering the pore size, pore size distribution, and pore stability that dictate the transport behaviour of the membranes.[20]

In order to generate porosity inside polyelectrolyte membranes, there have been several strategies developed, such as layer-by-layer deposition, templating, *etc*. It is also possible to take advantage of interpolyelectrolyte complexation between two oppositely charged polyelectrolytes (or a polyelectrolyte and an oppositely charged species) to construct porous polyelectrolyte membranes if a phase-separation process can arise simultaneously.[8,21] Our group previously exploited this complexation technique to create porous membranes based on two components, (*i.e.* an imidazolium poly(ionic liquid) (PIL), which is a polyelectrolyte built up from ionic liquid (IL) monomers,[22] and a multiacid compound that is usually organic compounds with multiple carboxylic acid units or poly(acrylic acid)).[23-25] These PIL membranes are versatile because by changing the IL moiety, its polymer characteristics, or the multiacid type that electrostatically cross-links the PIL, it is possible to confer different pore size to target different separation or transport applications.[25-29] Nevertheless, the first generation of porous PIL membranes suffered from a stability issue, as the interpolyelectrolyte complex membrane is ionic in nature and in a highly ionic environment undergoes partial, if not full, dissociation. This instability issue restricts the porous PIL membranes to be used in the absence of liquid electrolytes either in aqueous or organic solution. To address this issue, an easy strategy is to introduce covalent cross-links in addition to the existing electrostatic, noncovalent ones. This concept has already been exploited in covalently cross-linked amino- and carboxylat-containing polymer chains, forming an amide linkage for a better control over the swelling properties of the resulting membranes.[30-32] Taking advantage of the "click-"chemistry represents another way to covalently cross-link membranes forming a 1,2,3-triazole ring.[33] Moreover, diols have been proven to be able to covalently cross-link carboxylate groups upon ester formation.[34]

In this study, we describe a post-synthetic approach to introduce extra covalent cross-links into porous PIL membranes that are purely electrostatically cross-linked. These added covalent cross-links are able to improve membrane stability towards concentrated aqueous electrolyte solutions. The electrostatically cross-linked porous PIL membranes were obtained in an identical way to our previous work using a nitrile-containing imidazolium PIL and poly(acrylic acid) (PAA),[25-27] except that a dicyanamide (DCA) anion instead of bis(trifluoromethane sulfonyl)imide was used. It is well-known that at elevated temperatures, the nitrile groups or dicyanamide anion can undergo cyclisation reactions to produce a 1,3,5-triazine network.[35-40] Therefore, we exploited this reaction to create the covalent cross-links inside the prefabricated membrane.

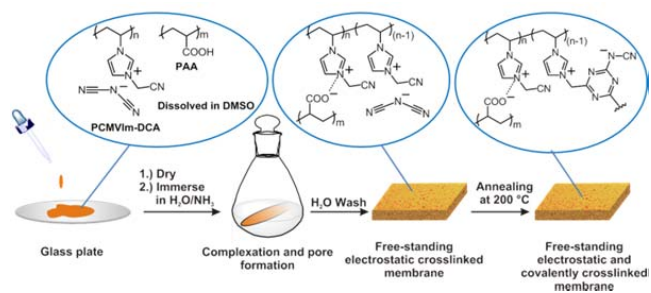

**Scheme 1.** The most relevant steps in the synthesis of covalently cross-linked imidazolium PIL membranes. The insets show the proposed chemical structure of the membranes in each step.

The overall procedure to synthesize the covalently cross-linked membrane is briefly shown in Scheme 1. The electrostatically cross-linked membrane was obtained via a procedure detailed in our previous report but employed a PIL with a different anion, poly(3-cyanomethyl-1-vinylimidazolium dicyanamide) (PCMVIm-DCA).[25] In order to trigger the cyclization reaction between the nitrile groups along the polymer backbone and in the DCA anions, the prefabricated membrane was thermally annealed at 200 °C for 2 hours in an $N_2$ atmosphere. It is worth mentioning that the resulting covalent cross-links do not replace but rather penetrate into the electrostatic ones. A sketch of the proposed chemical structure of the final electrostatic and covalently cross-linked PIL membrane is shown in Scheme 1.

The pore morphology before and after annealing was investigated by scanning electron microscopy (SEM). The average thickness of the membranes was 50 µm. As shown in Figure 1, the pores of sub-micrometer size are unaltered after membrane thermal cross-linking, providing evidence for the perfect preservation of the membrane porous structure after the triazine network formation. The average pore size, calculated from statistical analysis of the SEM images, is 300 ±100 nm before the annealing and 300 ±100 nm after. Given the experimental error in the SEM-based pore size imaging, these two values are no different from each other, and the annealed membrane has a porous structure identical to the native one. Moreover, the pore size distribution curves, before and after annealing, are practically the same (see Figure S1 in the supporting information).

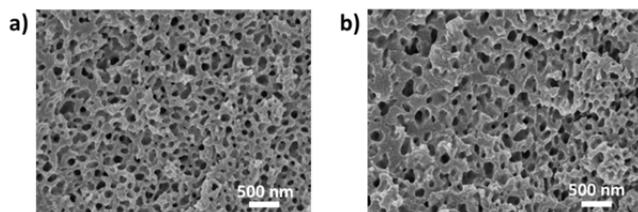

**Figure 1.** SEM images of the cross-section of the membrane before (a) and after (b) annealing.

To analyse the process of the thermal covalent cross-linking via the PIL chain pendant nitrile groups and DCA anions, the thermogravimetric profiles of the pristine porous PIL membrane, and its two structural components (*i.e.* PAA and PCMVIm-DCA) were collected and are shown in Figure S2 of the supporting information. PCMVIm-DCA is thermally more stable than PAA, showing a 2 wt% mass loss at 306 °C and 10 wt% at 335 °C. In comparison, PAA lost 2 wt% of its mass at 192 °C, and 10 wt% at 232 °C. Nevertheless, when PAA is in its dissociated form and electrostatically cross-linked with PCMVIm-DCA in the membrane, its 2 wt% mass loss was observed at 230 °C, and 10 wt% weight loss at 262 °C. This increase in thermal stability for PAA due to ionization allows annealing of the membrane at 200 °C without side reactions (*i.e.* the cyano/nitrile cyclization reaction is the only event). The thermal stability of the pristine porous PIL membrane was then compared to the one after the covalent crosslinking. Both thermogravimetric profiles in Figure S3 (see the Supporting Information) appear identical and overlapping. This high similarity is reasonably understandable, as the trimerization reaction at 200 °C does not produce any structural fragmentation that leaves the system.

The electrostatically cross-linked membrane was monitored by means of FTIR spectroscopy during an annealing process over a broad temperature range from 25 °C to 260 °C in an $N_2$ atmosphere. The collected FTIR spectra are shown in Figure 2. The starting spectrum at 25 °C (Figure 2a) is used as reference and displays the vibrational features of PCMVIm-DCA overlapped with those of deprotonated PAA. The first noticeable variation in the spectrum was observed once the temperature of the membrane reaches 195 °C (Figure 2b). The erosion of the bands at 2198 $cm^{-1}$ and 2143 $cm^{-1}$, respectively, (related to $\nu_s(C\equiv N)$ and $\nu_{as}(C\equiv N)$ of the dicyanamide anion) and the erosion of the band at 2237 $cm^{-1}$ (related to $\nu(C\equiv N)$ of the dangling nitrile group) is evident.[41] These bands, though much reduced in intensity, are partially retained after the annealing process since not all nitrile moieties reacted and ended up in triazine rings. Simultaneously, some new bands arise that can be assigned to the newly formed triazine rings. For instance, the band at 1685 $cm^{-1}$ is ascribable to the $\nu$ (C=N), while the two bands at 1225 $cm^{-1}$ and 1080 $cm^{-1}$ to $\nu_{as}$(C-N) and $\nu_s$(C-N), respectively.[42,43] These spectroscopic data support the formation of a triazine ring between the dicyanamide anion and nitrile moiety present in PCMVIm-DCA. The same behavior is evident for the reference annealing experiment conducted with pristine PCMVIm-DCA (see Figures S4-5 in the supporting information). The deprotonated PAA, as one part of the membrane component and to some extent the anion of the cationic imidazolium polymer, is not affected in the triazine cyclization cross-linking reaction. PAA stability after the annealing was confirmed by the unaltered symmetric and antisymmetric carboxylic group stretching at 1397 $cm^{-1}$ and 1572 $cm^{-1}$, respectively.[43] Raising the temperature to 230 °C (Figure 2c) and 250 °C (Figure 2d), the transformations of the spectral features are still clearly visible. Only at 250 °C does a new carbonyl stretching band appear at 1708 $cm^{-1}$, probably ascribable to a partial decomposition of the membrane. The aforementioned FTIR experiments revealed the cross-linking onset temperature to be 195 °C. Therefore, the annealing condition for the pristine membranes was chosen to be 200 °C for 2 h. The FTIR spectrum after this annealing treatment is shown in Figure 2e., in which the dicyanamide bands at 2198 $cm^{-1}$ and 2143 $cm^{-1}$ shrink dramatically, indicating a high degree of cyclization reaction in the membrane. Comparing it with the spectra collected at different temperatures, it is evident that the chosen annealing conditions generate the covalent cross-linking without degrading the membrane chemical structure.

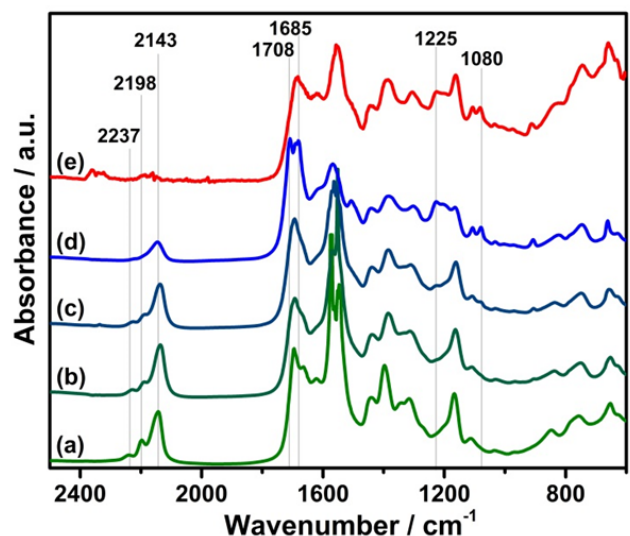

**Figure 2.** *In-situ* FTIR spectra of the initial electrostatically cross-linked membranes of PCMVIm-DCA and PAA. The spectra were collected at sample temperatures of 25 °C (a), 195 °C (b), 230 °C (c), 250 °C (d), respectively. For comparison, the FTIR spectrum of the membrane annealed at 200 °C for 2 hours in N₂ atmosphere is shown (e). Vertical lines indicate the relevant FTIR wavenumbers discussed in the manuscript.

Since the membrane is an insoluble, covalently cross-linked solid, 13C cross-polarization magic angle spinning (CP-MAS) solid-state NMR spectroscopy was performed to extract structural information about the generation of the covalent cross-links in the annealing step. The CPMAS NMR spectrum of the initial electrostatically cross-linked membrane is shown in Figure 3a with the corresponding signal assignations as shown in the scheme. The spectrum shows expected signals related to PCMVIm-DCA and PAA. Between 40 and 53 ppm, there are the alkyl $^{13}$C signals (**a,b,c,i** and **j**). The $^{13}$C signals at 123 ppm (**e** and **f**) and 136 ppm (**d**) can be assigned to the carbons 4, 5 and 6 of the imidazolium ring, respectively. The two $^{13}$C signals at 182 ppm (**k**) and 168 ppm (**l**) stem from the PAA carboxyl groups, in the deprotonated and protonated forms, respectively. The $^{13}$C signals of the nitrile carbon fall at 119 ppm (**g** and **h**) and are present as a shoulder of the imidazolium carbon band. These signal assignations were made by comparison of the solution NMR spectra of the two polymers and sodium dicyanamide reported in Figure S6-8. After membrane annealing, changes in the $^{13}$C CP-MAS NMR spectrum are evident (Figure 3b). First of all, two new $^{13}$C signals (**o** and **n**) arise, respectively, at 128 and 123 ppm. The first one (**o**) is from the nitrile moiety directly linked to the negatively charged nitrogen, connected to 1,3,5-triazine ring.[44] The second one (**n**) is from the carbons of 1,3,5-triazine, proving the successful closure of the aromatic ring.[45-47] The signals related to the nitrile carbon at 119 ppm (**g** and **h**) decrease after membrane annealing and their presence as a shoulder band is not clearly visible anymore. The signal of some protonated carboxylic groups at 168 ppm (**l**) was unaltered after the annealing, and the signal of the de-protonated carboxylic group at 182 ppm (**k**) remained as a shoulder of the 1,3,5-triazine carbon signal (**n**), supporting once again PAA stability during the annealing step. All these changes in the $^{13}$C CP-MAS NMR signals clearly confirm the formation of 1,3,5-triazine rings from the PIL nitrile groups and DCA anions and to generate new covalent bonds after the annealing step.

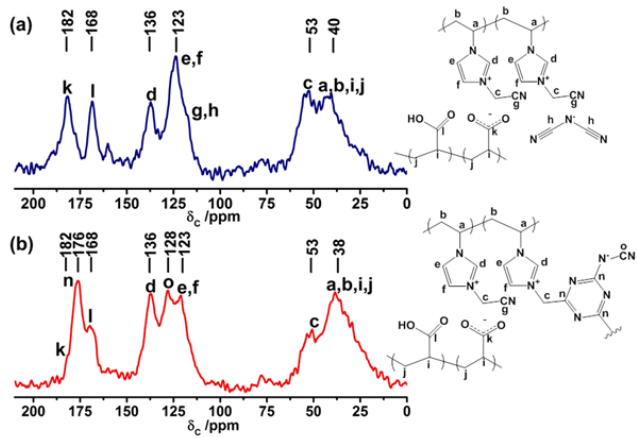

**Figure 3.** $^{13}$C CP-MAS solid-state NMR spectra of the initially electrostatically cross-linked membrane before covalent cross-linking (a) and of the covalently cross-linked one (b). Membrane structures with $^{13}$C signals assignations are reported on the right side.

A key parameter for membrane stability is retention of pore structure in a highly ionic environment; therefore, we tested the stability of the membranes in three electrolyte solutions. LiTf₂N and KPF₆, both in water and LiPF₆ in a 1:1 volumetric mixture of ethylene carbonate and dimethyl carbonate (a common electrolyte for Li-ion batteries) at 1 M concentration, were chosen as model environments to test the solution stabilities of the pristine porous PIL membrane and the annealed one for 3 days. The pore structure of the membrane before and after treatment was studied by SEM analysis (see Figure S9 in the supporting information). Before the annealing process, the membrane was not stable in ionic solutions because high ionic strength can dissociate the electrostatic cross-linking in the membrane and degrade it (i.e., change pore size distribution). As analyzed, before the salt treatment the initial electrostatically cross-linked membrane showed a narrow pore size distribution with a mean pore diameter of 120 ± 60 nm, while after treatment with an aqueous LiTf₂N solution, the average pore size increased to 500 ± 300 nm (Figure 4a and Table S1) with a rather broad size distribution profile. In contrast, the subsequently thermally covalently cross-linked membrane well retains its pore morphology in all the tested ionic solutions, as summarized in Table S1. When tested in the aqueous system, the LiTf₂N solution drops the average pore size slightly from 170 ± 20 nm to 140 ± 10 nm, a decrease of ~20% in average size (Figure 4b). This effect might result from anion exchange between Tf₂N⁻ and the residual DCA⁻ anions. In the case of the KPF₆ solution, the pore size slightly decreases from 300 ± 100 nm to 260 ± 100 nm by ~12% (Figure 4c). In the stability test in organic media, here a LiPF₆ solution in ethylene carbonate - dimethyl carbonate (v/v ~ 1:1), neither the average pore size nor the pore size distribution change after salt treatment (Figure 4d).

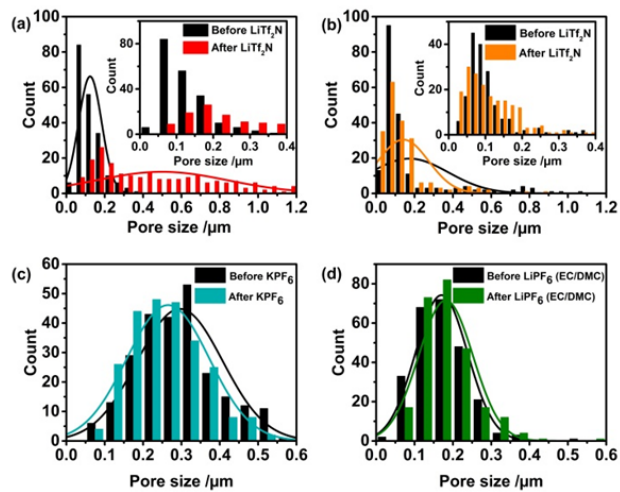

**Figure 4.** a) The initial non-annealed membrane (containing only electrostatic crosslinks) before (blue) and after (red) treatment with LiTf₂N solution in water; b)-d) Pore size distribution curves of annealed membranes before (black) and after salt treatment (colored): b) LiTf₂N in water; c) KPF₆ in water; d) LiPF₆ in ethylene carbonate/dimethyl carbonate (1:1 v/v).

In summary, electrostatically cross-linked porous membranes built up *via* interpolyelectrolyte complexation were stabilized by *in-situ* formation of an additional covalent cross-link network. The covalent bond is generated upon a solid-state cyclization reaction of dicyanamide anion with the dangling nitrile moieties. This covalent cross-linking has been proven by an extensive characterization of the membrane before and after annealing treatment at 200 °C. Even if the chemical structure is different because of the additional covalent cross-links, the pore structure of the membrane is retained after annealing. The covalently cross-linked membrane showed improved stability in various salt solutions, opening up the opportunity to study this membrane for applications requiring highly ionic solution environments, such as separators in Li-ion batteries.

## ASSOCIATED CONTENT

### Supporting Information

Full experimental information, thermogravimetric data, ATR-FTIR data, $^{13}$C CP-MAS NMR spectra of the starting polymers and average pore size data.

## AUTHOR INFORMATION

† K. Täuber and A. Dani contributed equally to this work

**Corresponding Author**

Jiayin Yuan (Email: jiayin.yuan@mpikg.mpg.de)

**FUNDING SOURCES**

This work was supported by the Max Planck Society, Germany an ERC (European Research Council) Starting Grant (project number 639720 – NAPOLI)

**ACKNOWLEDGMENTS**

Thanks are due to Dr. habil. Bernhard Schartel and Patrick Klack from Bundesanstalt für Materialforschung und -prüfung (BAM) for the FTIR measurement at variable temperature. Thanks are due to Dr. Jan Dirk Epping from TU Berlin for the SS-NMR analyses.

# Covalent Cross-linking of Porous Poly(Ionic Liquid) Membranes via a Triazine Network

Karoline Täuber,[†] Alessandro Dani,[†] Jiayin Yuan*

Department of Colloid Chemistry, Max Planck Institute of Colloids and Interfaces, Am Mühlenberg 1 OT Golm, D-14476 Potsdam, Germany.

**Materials and characterization methods**

Vinylimidazole, bromoacetonitrile, AIBN, sodium dicyanamide, poly(acrylic acid) (Mw: 450 kDa) and ammonium hydroxide solution 28% in water were obtained from Sigma-Aldrich and used without further purification. Dimethyl sulfoxide (DMSO) was of analytical grade.

Scanning electron microscopy (SEM) was performed using a GEMINI LEO 1550 microscope at 3 kV, and samples were coated with gold before examination. ATR-FTIR spectra were collected with a Nicolet iS5 FT-IR instrument from ThermoFisher Scientific equipped with a single-reflection ATR diamond with a working range from 4000 $cm^{-1}$ to 500 $cm^{-1}$ with 64 scans per spectrum and 128 scans for the background with a resolution of 2 $cm^{-1}$. Thermogravimetric measurements were performed with a TG 209 F1 Libra thermomicrobalance from Netzsch. The samples were analyzed in a platinum crucible at a heating rate of 10 K $min^{-1}$ under a $N_2$ flow of 30 mL $min^{-1}$. The FT-IR set-up used to study the cross-linking mechanism of the membranes consisted of a FT-IR 600 Linkam stage cell connected to a Bruker Vertex 70 FT-IR spectrometer. The membrane is placed in the FTIR cell and is heated at a rate of 5 K $min^{-1}$ from 25 °C to 260 °C under nitrogen flow (150 mL $min^{-1}$). FT-IR spectra are recorded every 10 s during the heating of the sample. $^1H$ NMR and $^{13}C\{^1H\}$ NMR spectra were collected using a Bruker DPX-400 spectrometer in $D_2O$ as solvent. Solid-state NMR spectra were recorded with a Bruker Avance 400 MHz spectrometer operating at 100.56 MHz for $^{13}C$ and 399.88 MHz for $^1H$ and $^1H$-$^{13}C$ cross-polarization magic angle spinning (CP-MAS) NMR experiments were carried out at a MAS rate of 10 kHz using a 4 mm diameter MAS HX double-resonance probe. The $^1H$ π/2 pulse length was 3.1 µs and two pulse phase modulation (TPPM) heteronuclear dipolar decoupling was used during acquisition. The spectra were measured using contact time of 2.0 ms for $^{13}C$ and recycle delays of 2 s. All $^{13}C$ spectra are referenced to external TMS at 0 ppm using adamantane as a secondary reference. Pore size distribution of the membrane was analyzed on membrane cross-section SEM images by means of Adobe Photoshop CS5 software. All the data were elaborated using OriginPro2015 software.

**Synthesis of poly(3-cyanomethyl-1-vinylimidazoium dicyanamide)**

Poly(3-cyanomethyl-1-vinylimidazoium bromide) was prepared according to our previous method (J. Am. Chem. Soc. 2012, 134, 11852–11855). Afterwards, this polymer undergoes anion exchange towards dicyanamide following this procedure: a solution of sodium dicyanamide (0.917 g, 10.3 mmol) in water (30,0 mL) was slowly dropped on a solution of poly(3-cyanomethyl-1-vinylimidazoium bromide) (2 g) in water (60,0 mL). The obtained poly(3-cyanomethyl-1-vinylimidazoium dicyanamide) (PCMVIm-DCA) precipitated in water due to the presence of excess charge in solution.

The PIL was filtered off and dried until constant weight. The polymer yield of the anion exchange is 92,7%. The EDX shows a bromine content of 34.8 wt% in poly(3-cyanomethyl-1-vinylimidazoium bromide) and of 13.5 wt% for poly(3-cyanomethyl-1-vinylimidazoium dicyanamide); therefore, it is possible to calculate an anion exchange rate of 61 mol%. For the purpose of membrane fabrication, this anion exchange amount has proven to be sufficiently high.

**Synthesis of initial electrostatically cross-linked PIL membrane**

A homogeneous solution of poly(3-cyanomethyl-1-vinylimidazoium dicyanamide) (0.200 g, 1 mmol of repeating unit) and poly(acrylic acid) (PAA) (0.072 g, 1 mmol of repeating unit) in DMSO (2 mL) is cast onto a glass plate and the solvent is evaporated at 80 °C. The obtained dry PCMVIm-DCA/PAA blend film, coated on the glass plate, is immersed in an aqueous $NH_3$ solution (~0.2 % w/w, 10 mL) for 2 h. During this period, ammonia solution slowly penetrated into the polymers film, neutralizes PAA into the charged poly(ammonium acrylate) salt and activates the electrostatic cross-linking between PIL and poly(ammonium acrylate). It should be noticed that the inter-polyelectrolyte complexation is an entropy-driven process, in which the complexation proceeds only to a certain extent. Thus, the DCA anions remain in a significant amount in the PIL. After two hours, the electrostatically cross-linked membrane was washed with water in order to remove the excess of $NH_4OH$ and the formed $NH_4DCA$, and it was then dried in a vacuum oven for 2 hours.

**Synthesis of electrostatically and covalently cross-linked membrane**

The obtained initial electrostatic cross-linked membrane was annealed at 200 °C for 2 h under a nitrogen atmosphere, in order to form the additional covalent cross-links. The resulting electrostatically and covalent cross-linked membrane was used without any further clean-up or drying.

**Characterization Data**

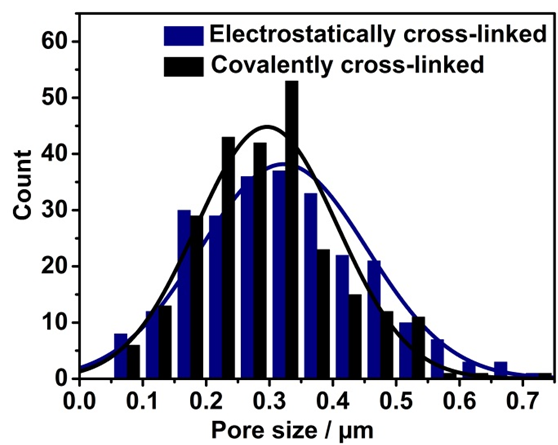

**Figure S1.** Pore size distribution of porous membrane before (blue columns) and after annealing (black columns).

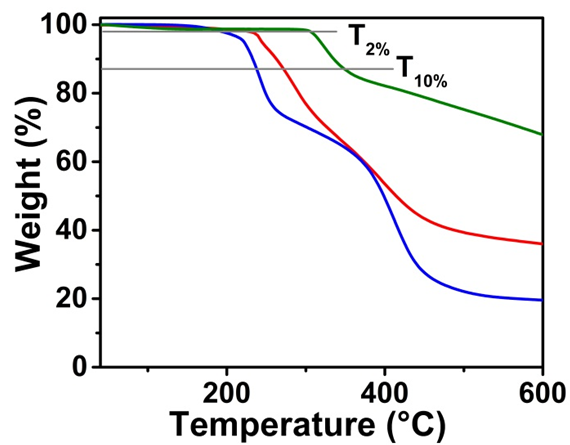

**Figure S2.** Thermogravimetric profiles of: PAA (blue curve), PCMVIm-DCA (green curve) and of the membrane originating from their electrostatic complexation (red curve).

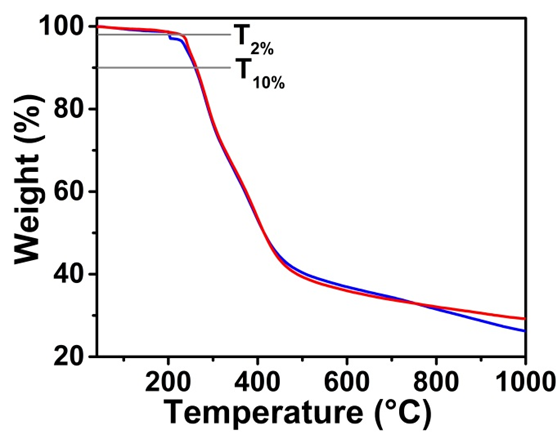

**Figure S3.** Thermogravimetric profiles of electrostatic cross-linked membrane (red curve) compared with the one of covalently cross-linked membrane (blue curve).

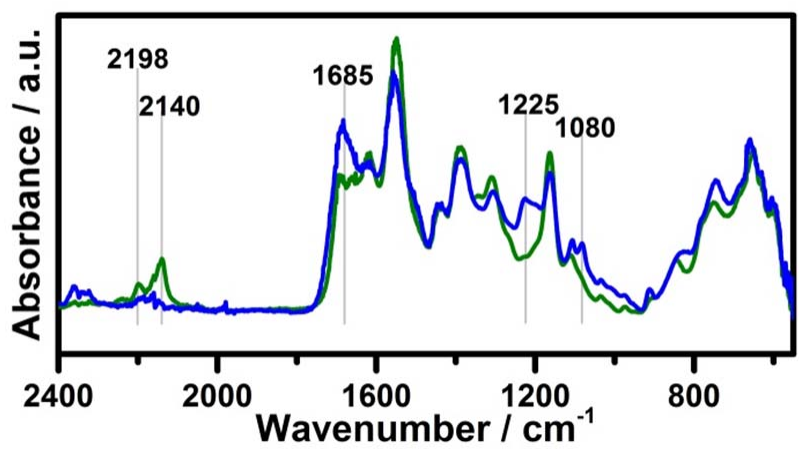

**Figure S4.** FTIR-ATR spectra of PCMVIm-DCA before (green curve) and after annealing process at 200°C (blue curve).

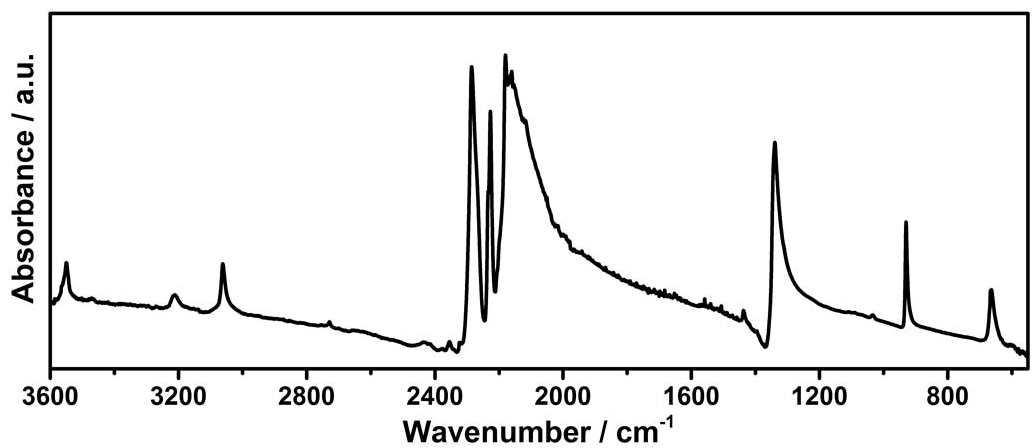

**Figure S5.** ATR-FTIR spectrum of sodium dicyanamide.

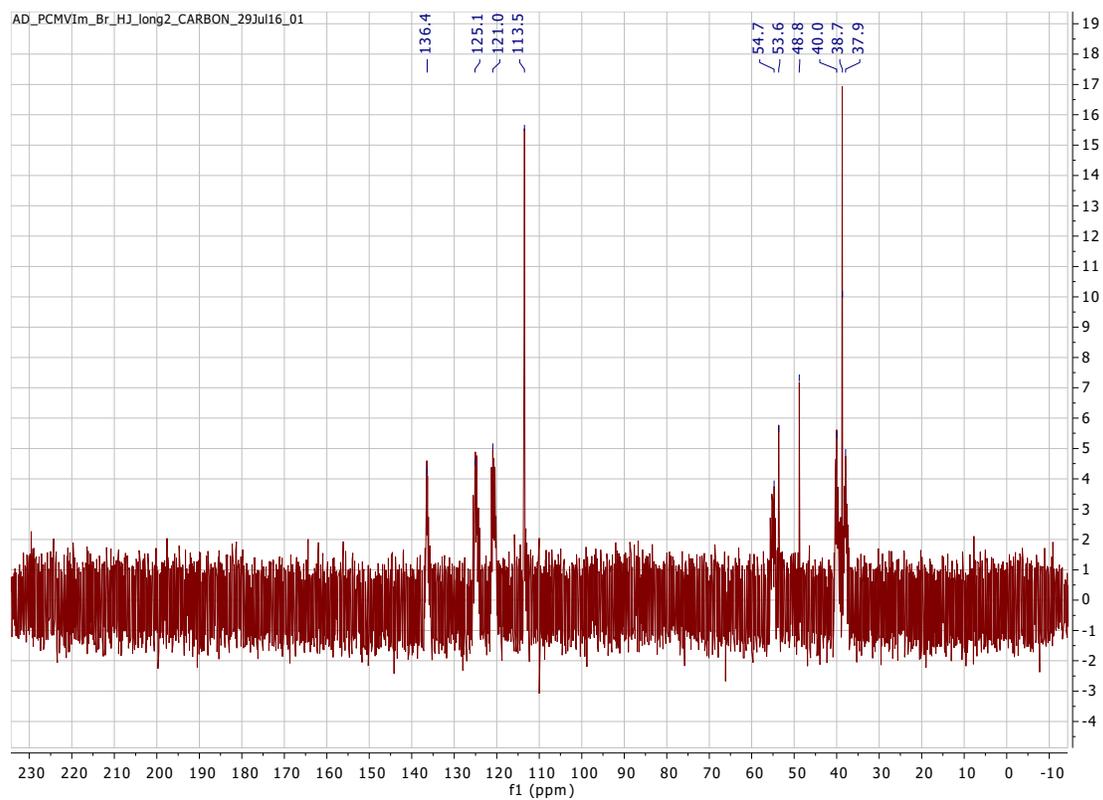

**Figure S6**. $^{13}$C-NMR spectra of Poly(CMVIm-Br) in D$_2$O

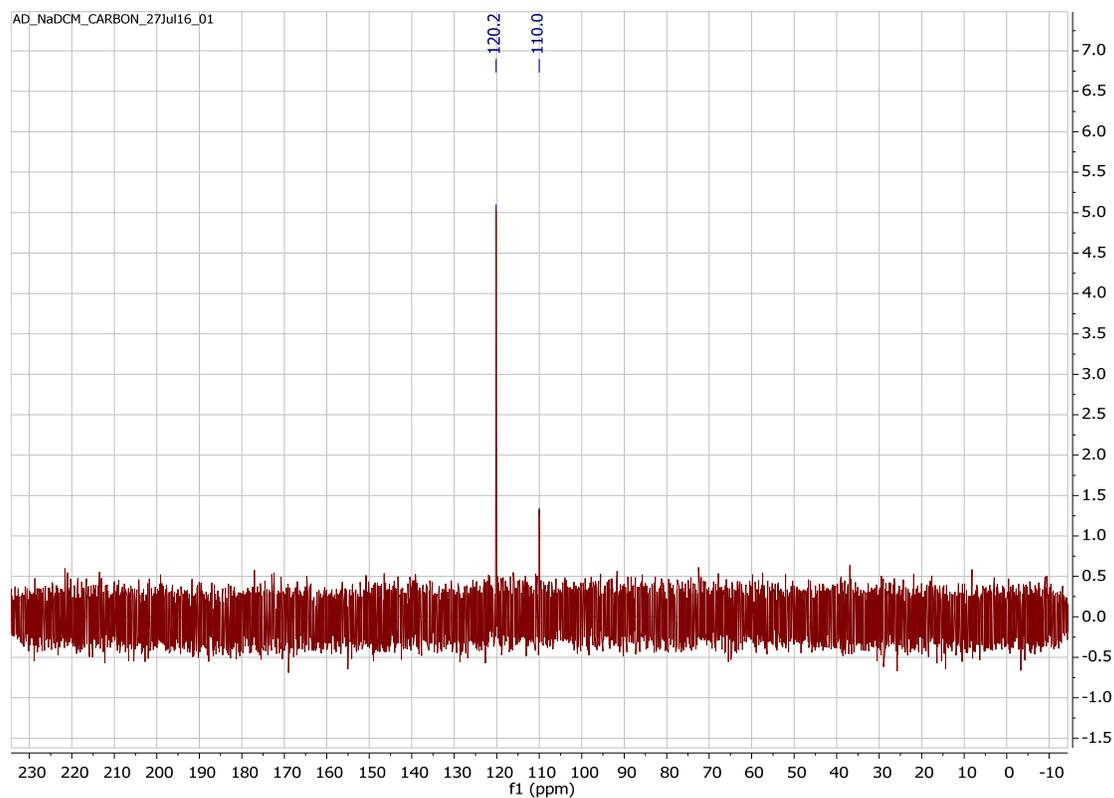

**Figure S7.** $^{13}$C-NMR spectra of sodium dicyanamide in D$_2$O

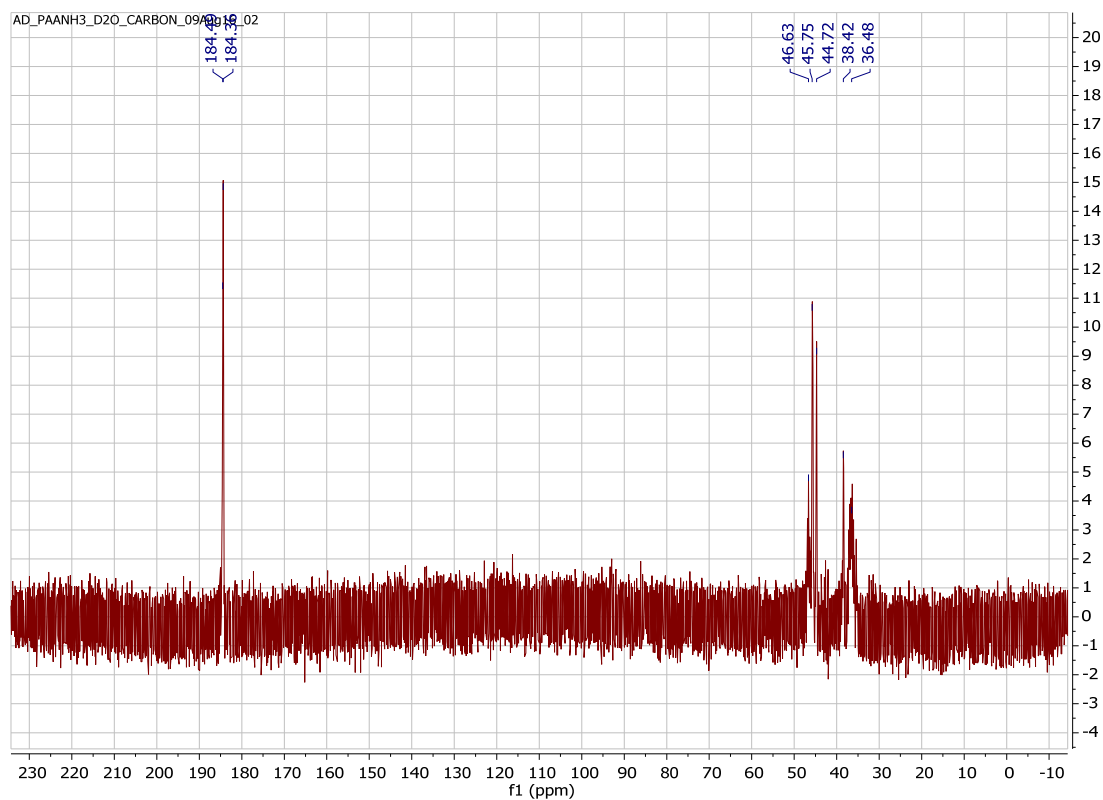

**Figure S8.** $^{13}$C-NMR spectra of poly(acrylic acid) ammonium salt solution in D$_2$O.

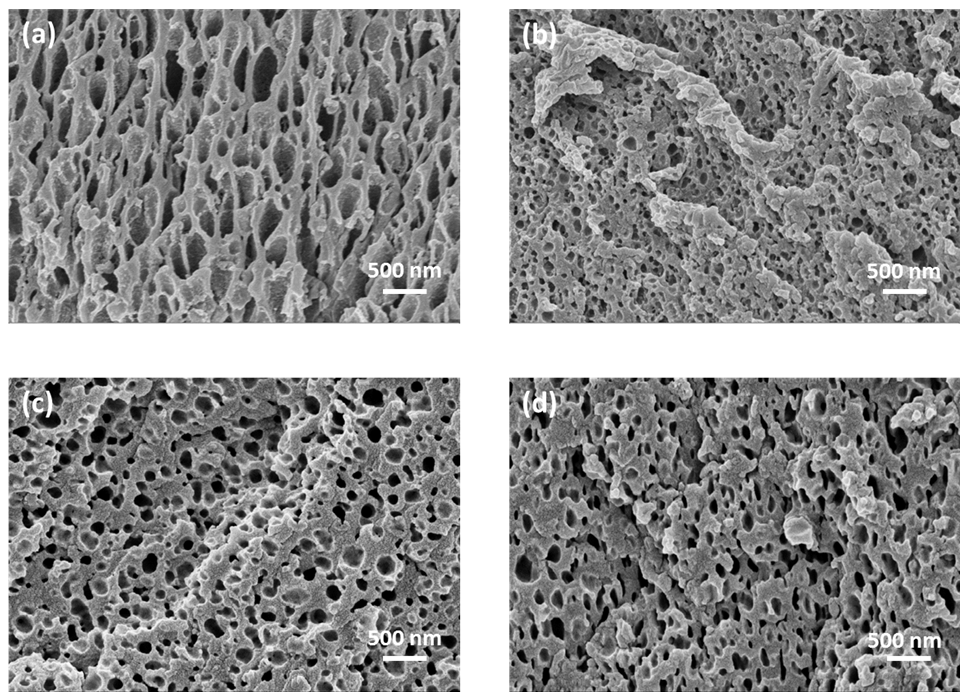

**Figure S9.** a) SEM images of the initial non-annealed membrane after treatment with LiTf$_2$N solution in water; b)-d) SEM images of annealed membranes after salt treatment: b) LiTf$_2$N in water; c) KPF$_6$ in water; d) LiPF$_6$ in ethylene carbonate/dimethyl carbonate (1:1 v/v).

**Table S1.** Average pore size distribution and standard deviation of electrostatical and covalent cross-linked membranes before and after treatment with salt solution.

| membrane | treatment | average pore size nm | standard deviation nm |
|---|---|---|---|
| Initial electrostatically cross-linked membrane | Before LiTf$_2$N solution soaking | 120 | 60 |
| | After LiTf$_2$N solution soaking | 500 | 300 |
| Electrostatically and covalently cross-linked membrane | Before LiTf$_2$N solution soaking | 170 | 20 |
| | After LiTf$_2$N solution soaking | 140 | 10 |
| | Before KPF$_6$ solution soaking | 300 | 100 |
| | After KPF$_6$ solution soaking | 260 | 100 |
| | Before LiPF$_6$ solution soaking | 170 | 70 |
| | After LiPF$_6$ solution soaking | 180 | 70 |